\documentclass[10pt,aps,twocolumn,floatfix,tightenlines,showpacs,showkeys,preprintnumbers,nofootinbib, longbibliography]{revtex4-1}

\pdfoutput=1
\usepackage[colorlinks=true,breaklinks=true]{hyperref}
\hypersetup{allcolors=[rgb]{0.0 0.0 0.6},linkcolor=[rgb]{0.8 0.0 0.4}}
\usepackage{amsmath,amssymb,mathtools}
\usepackage{epsfig}  
\usepackage{graphicx}   
\usepackage{commath}
\usepackage{slashed}             
\usepackage{url}
\usepackage{color}
\usepackage{multirow}
\usepackage{placeins}
\usepackage[dvipsnames]{xcolor}
\allowdisplaybreaks

\bibpunct{[}{]}{,}{n}{}{,}

\newcommand{\comments}[1]{}
\newcommand{\dsum}{\displaystyle \sum}
\newcommand{\Tl}{\mathcal{T}_\ell}
\newcommand{\Sl}{\mathcal{S}_\ell}
\newcommand{\Psil}{\Psi_\ell}
\newcommand{\Jl}{\mathcal{J}_\ell}
\newcommand{\Nl}{\mathcal{N}_\ell}
\newcommand{\Kl}{\mathcal{K}_\ell}
\newcommand{\Bl}{\mathcal{B}_\ell}
\newcommand{\Hl}{\mathcal{H}_\ell}
\newcommand{\Fl}{\mathcal{F}_\ell}
\newcommand{\bS}{\boldsymbol S}

\begin{document}

\title{Self-Scattering of Multi-level Dark Matter and New Dissipation Mechanisms}
\title{New Dissipation Mechanisms from Multi-level Dark Matter Scattering}

\author{Anirban~Das}
\email{anirbandas@theory.tifr.res.in }              
\author{Basudeb~Dasgupta}
\email{bdasgupta@theory.tifr.res.in}
\affiliation{Tata Institute of Fundamental Research,
             Homi Bhabha Road, Mumbai, 400005, India.}

\preprint{TIFR/TH/17-32}
\date{February 6, 2018}

\begin{abstract}
Multi-level dark matter with diagonal and off-diagonal interactions shows a rich phenomenology in its self-scattering. If the interactions are mediated by a particle that is less massive than the dark matter, Sommerfeld effect can lead to resonant enhancement of the scattering. For mediators lighter than the level separation, dark matter particles can upscatter to excited states and de-excite by emitting these mediators. We compute these cross-sections, both above and below the kinematic threshold, in a generic two-component dark matter model and identify the large inelastic cross-section as a result of maximal mixing between the two states. A new route for cooling of large dark matter halos and a new drag force between two colliding halos are identified and shown to arise purely from the inelastic scattering.

\end{abstract}

\maketitle

\section{\label{sec:intro}Introduction}
Galaxies are observed to be surrounded by more massive halo-like structures made of a substance whose particle nature still remains unknown. The formation of these halos is predicted within the paradigm that dark matter (DM) is made of cold collisionless particles. Agreement with several other cosmological and astrophysical observations at widely different length scales has given strong support for this simple paradigm. Notwithstanding this success, it fails to provide explanation for some discrepancies between the predictions and observations of the shapes and abundances of these halos at sub-galactic length scales, {viz.}, the \textit{small scale problems}\,\cite{DelPopolo:2016emo, Tulin:2017ara}.

One of these problems is often referred to as the \textit{core-cusp problem}\,\cite{Moore:1994, Flores:1994gz, Navarro:1996gj, Moore:1999gc}. The central density profile of dwarf galaxies is observed to be cored, while simulations with the standard cold collisionless DM typically lead to a denser cuspy profile ($\sim 1/r$) near the centre. Complex astrophysical processes involving baryonic matter, e.g., tidal effects and supernova explosions remove matter from the central region, may lead to cored profiles reducing the discrepancy\,\cite{ElZant:2001re, ElZant:2003rp, Mashchenko:2007jp, 2010MNRAS.406.1290P, 2010Natur.463..203G, 2011AJ....142...24O, Oh:2015xoa}. Indeed, these feedback processes could very well be the missing ingredient in the simulations. However, it is challenging to accurately model these processes and it is not yet established if one can obtain sufficient feedback in realistic models\,\cite{Oman:2015xda, Sawala:2014xka, Marinacci:2013mha, Choi:2014hda}. As such, this problem remains open and motivates us to consider other possibilities as well.

Self-scattering of DM particles have been shown to be effective in solving the core-cusp problem\,\cite{Spergel:1999mh, Burkert:2000di}. In this class of models, known as self-interacting DM (SIDM), the DM particles have strong interactions with each other. During the non-linear phase of structure formation, when the central density of a halo becomes large,  the self-scattering generates outward pressure. When this pressure equates the gravitational pull on the matter, any further accumulation of DM at the centre ceases and a stable core is formed. While these interactions may be obtained with massive mediators, the cross section in such models is velocity-independent and is strongly constrained from various observations\,\cite{Buckley:2009in}. Simulations indicate that a hard-sphere scattering cross-section per unit DM mass $\sigma/M \sim 0.5-5\ {\rm cm}^2/$g  is required to  form the cores at the centres of galaxies with DM velocity $v_{\rm rms} = 30-100$ km/s\,\cite{Zavala:2012us, Elbert:2014bma, Kaplinghat:2015aga}. On the other hand, objects with larger $v_{\rm rms}$ put stronger bounds on these same cross-sections\,\cite{Markevitch:2003at, Clowe:2006eq, 2008ApJ...687..959B, Jee:2013gey, Jee:2014mca}. For example, the non-observation of drag force between the DM components of two merging clusters puts an upper bound $\sigma/M \lesssim 1\ {\rm cm}^2$/g\,\cite{Dawson:2011kf, Kim:2016ujt}. But a tighter constraint $\sigma/M\lesssim 0.1\ {\rm cm^2/g}$ is given by the stellar kinematics and weak lensing data in galaxy clusters\,\cite{Kaplinghat:2015aga}. As self-interaction helps virialize DM halos efficiently, they become rounder compared to the anisotropic halos predicted by collisionless DM scenario. Therefore the observed triaxialities of halos provides an upper bound to the strength of self-interaction among the DM particles. A comparison of observations with N-body simulation shows that a scattering cross-section as large as $\sigma/M \simeq 1\ {\rm cm^2/g}$ could be allowed at the larger velocity end, i.e., $v \sim 1000$ km/s\,\cite{Peter:2012jh}.

Therefore, one typically considers lighter mediators that lead to velocity-dependent cross-sections\,\cite{Feng:2009hw, Loeb:2010gj}. In these models, a velocity-dependent cross-section is obtained from a long-range interaction between DM particles, and have been shown to be able to satisfy observational constraints\,\cite{Peter:2012jh}. The long ranged interactions, on the other hand, have interesting phenomenology\,\cite{Ackerman:mha, Feng:2009mn}. When the mass of the mediator particle is smaller than that of the DM, \textit{Sommerfeld effect} may cause resonant scattering. In this regime, the cross-section is resonantly enhanced through virtual bound state formation\,\cite{Tulin:2012wi}.

A multi-component DM system brings new features in the scattering phenomenology and the dynamics of halo formation. In particular, it can solve the core-cusp problem by heat flow from the hotter outer region to the colder inner core. The possibility of elastic as well as inelastic scattering depending on the energy of the particles, gives a rich phenomenology that has not been explored fully. For example, the excitation and de-excitation of DM particles can give rise to observable indirect detection signals. Also, the energy dissipation from the inelastic scattering, followed by de-excitation, might lead to significant change in the shape and density profile of DM halos. Some of these features have been discussed in the context of atomic DM models in Ref.\,\cite{CyrRacine:2012fz, Cline:2013pca, Boddy:2016bbu}, two-level DM systems with purely off-diagonal interaction in Ref.\,\cite{Schutz:2014nka, Blennow:2016gde} and dark bremsstrahlung process\,\cite{Foot:2014uba}. Multi-level DM models also have interesting phenomenology in the context of direct and indirect detection experiments\,\cite{Dienes:2014via, Dienes:2017ylr, Das:2016ced}.

In this work, we take a two-level SIDM model with light particles mediating both diagonal and off-diagonal interactions. In this model, the DM particles can not only elastically scatter due to the diagonal interactions, but they may also get excited to the more massive partner of DM due to the off-diagonal interactions and subsequently de-excite by emitting the light mediator particles. This leads to additional dissipation. We compute these cross sections, analytically explain their behavior in various regimes, and study the cooling of DM halos due to the new dissipation mechanism. We further identify a new dissipation-induced \textit{drag force} between two colliding halos in such models.

The paper is organized as follows. In Sec.\,\ref{sec:formalism}, we discuss a formalism for two-level scattering with a description of the minimal SIDM model considered here, numerically compute the elastic and inelastic cross-sections, and explain their key features using simple analytical estimates. We then outline the key signatures and possible constraints in Sec.\,\ref{sec:result}, and, in Sec.\,\ref{sec:concl}, conclude with a summary of our results and avenues for future work.

\section{\label{sec:formalism}Two level DM \& scattering}
To discuss the phenomenology of a multi-level DM model, we concentrate on a simple two-level DM system with a small mass gap $\Delta$ between the two states, $\chi_1$ and $\chi_2$, with masses $M$ and $M+\Delta$, respectively. We further assume a dark $\mathcal{Z}_2$ symmetry under which $\chi_{1,2}$ have charges $\mp 1$, respectively. Two lighter scalars $\rho_1$ and $\rho_2$ with charges $\pm 1$, couple to the DM states as
\begin{equation}
 \label{eq:lag}
 \mathcal{L}_{\rm int} \supset f\rho_1\left(\bar{\chi}_1\chi_1-\bar{\chi}_2\chi_2\right)+f\rho_2\left(\bar{\chi}_1\chi_2+\bar{\chi}_2\chi_1\right)\,.
\end{equation}
For simplicity, we have assumed the masses and couplings of $\rho_{1,2}$ to be the same and equal to $m_\rho$ and $f$, respectively. Most of our discussion is insensitive to these simplifying assumptions, and we will outline how the results would change in a more general model, wherever necessary.

\begin{figure}[t]
 \begin{center}
  \includegraphics[width=0.93\columnwidth]{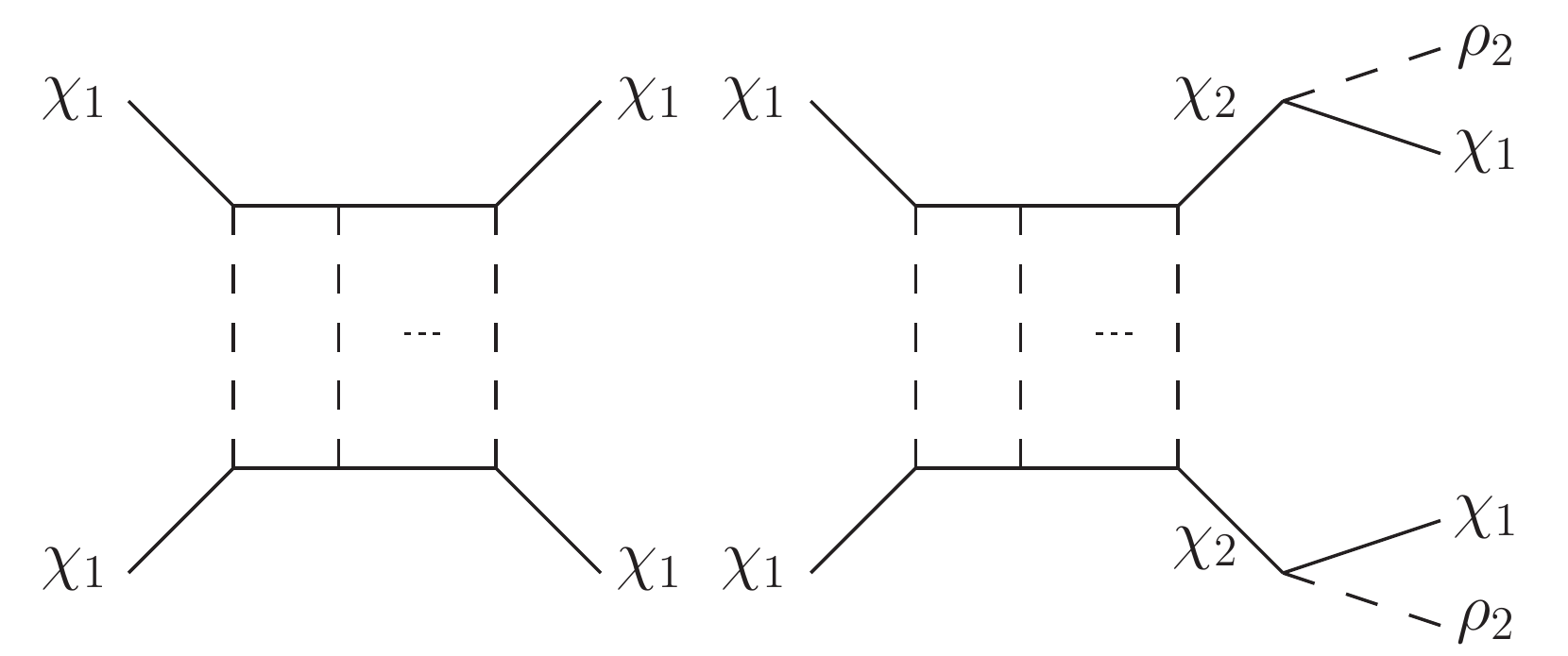}
  \caption{Typical Feynman diagrams for elastic self-scattering of DM in the ground state (left) and for upscattering induced decay (right). The intermediate vertical lines represent multiple exchanges of scalar $\rho$ particles in the nonrelativistic limit of the incoming DM particles.}
  \label{fig:scattering}
 \end{center}
\end{figure}

Two colliding DM particles that are initially in the ground state can either stay in the ground state (elastic) or upscatter to the excited state (inelastic). For upscattering to occur, the incoming particles need to have enough kinetic energy to overcome the mass gap $2\Delta$ between the two 2-body states. In addition, even if there is not enough kinetic energy, the excited state can be produced as virtual particles in the intermediate steps of the collisions. Moreover, the scattering cross-section between nonrelativistic DM particles is enhanced due to multiple exchanges of the light $\rho$ particle, influencing the Sommerfeld effect. Schematic Feynman diagrams for the possible elastic and inelastic scatterings are shown in Fig.\,\ref{fig:scattering} where the vertical lines represent exchange of many $\rho_{1,2}$ particles in the nonrelativistic regime. In the case of inelastic scattering, the final particles decay to the ground state by emitting two light particles. This process is essentially an energy loss process, and is expected to have interesting phenomenology. 

The scattering cross-sections are computed by calculating the transition amplitude between an allowed initial state $|i\rangle$ and final state $|f\rangle$. The possible 2-body states are $|\chi_1\chi_1\rangle,\ |\chi_2\chi_2\rangle$ and $|\chi_1\chi_2\rangle$. However, it is easy to see from Eq.(\ref{eq:lag}) that $|\chi_1\chi_1\rangle$ and $|\chi_2\chi_2\rangle$ are decoupled from $|\chi_1\chi_2\rangle$, due to the $\mathcal{Z}_2$ symmetry. Therefore, assuming that the DM particles are initially in the ground state, it suffices to work in a Hilbert space spanned by $|\chi_1\chi_1\rangle$ and $|\chi_2\chi_2\rangle$ only. One can remove this restriction, but it makes the calculation more difficult without yielding any qualitatively new features. This is our motivation for using two oppositely charged scalars, as opposed to a single scalar that one may naively think to be the simpler case. We neglect the scattering between $\chi_1$ and $\chi_2$, because $\chi_2$ decays soon after freeze-out and its abundance is rapidly depleted. For the same reason, the scattering between two $\chi_2$ particles is negligible. Therefore, we have two channels, i.e., $|\chi_1\chi_1\rangle \rightarrow |\chi_1\chi_1\rangle$ (elastic) and $|\chi_1\chi_1\rangle \rightarrow |\chi_2\chi_2\rangle$ (inelastic).

The overlap between two 2-body states is defined as $\Psi_{ab} \equiv \langle\chi_a\chi_a|\chi_b\chi_b\rangle$ with $a, b = 1,2$ and satisfies the Schr\"odinger equation,
\begin{equation}
 \label{eq:schro}
 \left[\dfrac{1}{r^2}\dfrac{d}{dr}\left(r^2\dfrac{d}{dr}\right)+k^2-\dfrac{\ell(\ell+1)}{r^2}-2\mu V(r)\right]\Psil(r) = 0\,,
\end{equation}
where $\ell$ is the orbital angular momentum, and $k$ and $\mu$ are two diagonal matrices with the momentum and reduced mass of the incoming 2-body state, respectively,
\begin{equation}
\label{eq:momentum}
 k = k_a\delta_{ab},\quad\rm{and}\quad \mu = \mu_\mathit{a} \delta_{\mathit{ab}}\,.
\end{equation}
The incoming momentum $k_a$ is different for the two 2-body states due to the presence of the mass gap $2\Delta = 2(M_2-M_1)$ between the states $|\chi_1\chi_1\rangle$ and $|\chi_2\chi_2\rangle$. Depending on the energy 
\begin{equation}
E_1=k_1^2/2\mu_1=\mu_1v^2/2
\end{equation}
of the incoming particles, two cases are possible:
\begin{itemize}
 \item \textit{Below threshold}, $\mu_1v^2/2<2\Delta$, when the initial energy of the incoming state is below threshold, then the heavier $|\chi_2\chi_2\rangle$ state is kinematically closed as $\chi_2$s cannot be produced onshell,
 \item \textit{Above threshold}, $\mu_1v^2/2>2\Delta$, when the incoming energy is above threshold, the excited state is open and DM particles can upscatter to the excited state.
\end{itemize}

The exchange of the scalar particles between the DM particles as dictated by the Lagrangian in Eq.(\ref{eq:lag}) gives rise to an attractive potential between the 2-body states in the nonrelativistic limit of the theory. The potential matrix $V(r)$ in Eq.(\ref{eq:schro}) is given by
\begin{equation}\label{eq:potential}
V = \begin{pmatrix}
V_1 & V_1\\
V_1 & V_1
\end{pmatrix} \quad {\rm with\ } V_1(r) = -\frac{\alpha\, e^{-m_\rho r}}{r},
\end{equation}
where $\alpha\equiv f^2/4\pi\,$. This attractive Yukawa potential matrix, with equal entries, is a result of assuming the same interaction strength between either pair of 2-body states. The structure of the potential matrix would be different in other DM models, e.g., a broken dark gauge symmetry would provide both attractive and repulsive interactions, which will become purely repulsive for a late-time asymmetric DM population. With additional scalars one can engineer diagonal and off-diagonal potentials of different strengths. As we shall see, the qualitative nature of many results discussed in this work remain unchanged as long as the matrix has nonvanishing off-diagonal components. Therefore we shall not delve into these details here.

The set of Schr\"odinger equations in Eq.(\ref{eq:schro}) is to be solved with appropriate boundary conditions for above and below threshold scatterings. The equations are solved for each partial wave $\ell$ and the large-$r$ wavefunctions are matched with plane wave solutions to extract the elements of the transition matrix $\Tl$ which consists of the transition amplitudes from state $|i\rangle$ to $|f\rangle$. The scattering matrix $\Sl$ is written as $\Sl\equiv \mathit{1}-\Tl$. Finally the scattering matrices are added upto some $\ell=\ell_{\rm max}$, to yield the total cross-section. Although theoretically $\ell_{\rm max}$ goes upto infinity, in practice a finite value must be chosen by ensuring numerical convergence of the sum. This value depends on the range of the potential and the momentum of the incoming particles. 

The total scattering cross-section $\sigma_{\rm tot}$ is given by
\begin{equation}
 \sigma_{\rm tot} = \int d\Omega\dfrac{d\sigma}{d\Omega}\,.
\end{equation}
This definition gives equal weight to all scattering angles which is useful in case of \emph{hard sphere scattering} mediated by a heavy particle with short range. For a small mediator mass, the cross-section is peaked in the forward and backward directions. 
While this leads to an overall large value of cross-section, the effective momentum transfer in each collision is small if the particles are identical. However, in a DM halo and in N-body computer simulations, momentum transfer during a collision is the quantity that determines the virialization and shape of a halo during its evolution and the dynamics of colliding halos. In Ref.\,\cite{PhysRevA.60.2118}, it was pointed out that the \emph{transfer cross-section} $\sigma_{\rm T}$, which removes the forward direction peak, is a more important quantity for transport phenomena. It is defined as
\begin{equation}\label{eq:transcs}
 \sigma_{\rm T} = \int d\Omega\dfrac{d\sigma}{d\Omega}(1-\cos\theta)\,.
\end{equation}
Another quantity which is often used is the \emph{viscosity cross-section} $\sigma_{\rm V}$\,\cite{PhysRevA.60.2118},
\begin{equation}\label{eq:viscositycs}
 \sigma_{\rm V} = \int d\Omega\dfrac{d\sigma}{d\Omega}\sin^2\theta\,.
\end{equation}
This definition removes the contributions of the backward scatterings, in addition. The details of the numerical calculations of these cross sections are given in appendix\,\ref{sec:appendix}. We now discuss the nature of the elastic and inelastic cross-sections in the limits when $\mu_1v^2/2 < 2\Delta$, i.e., below threshold and when $\mu_1v^2/2 > 2\Delta$, i.e., above threshold.

\subsection{Below Threshold}


\begin{figure}[!t]
\includegraphics[width=0.9\columnwidth]{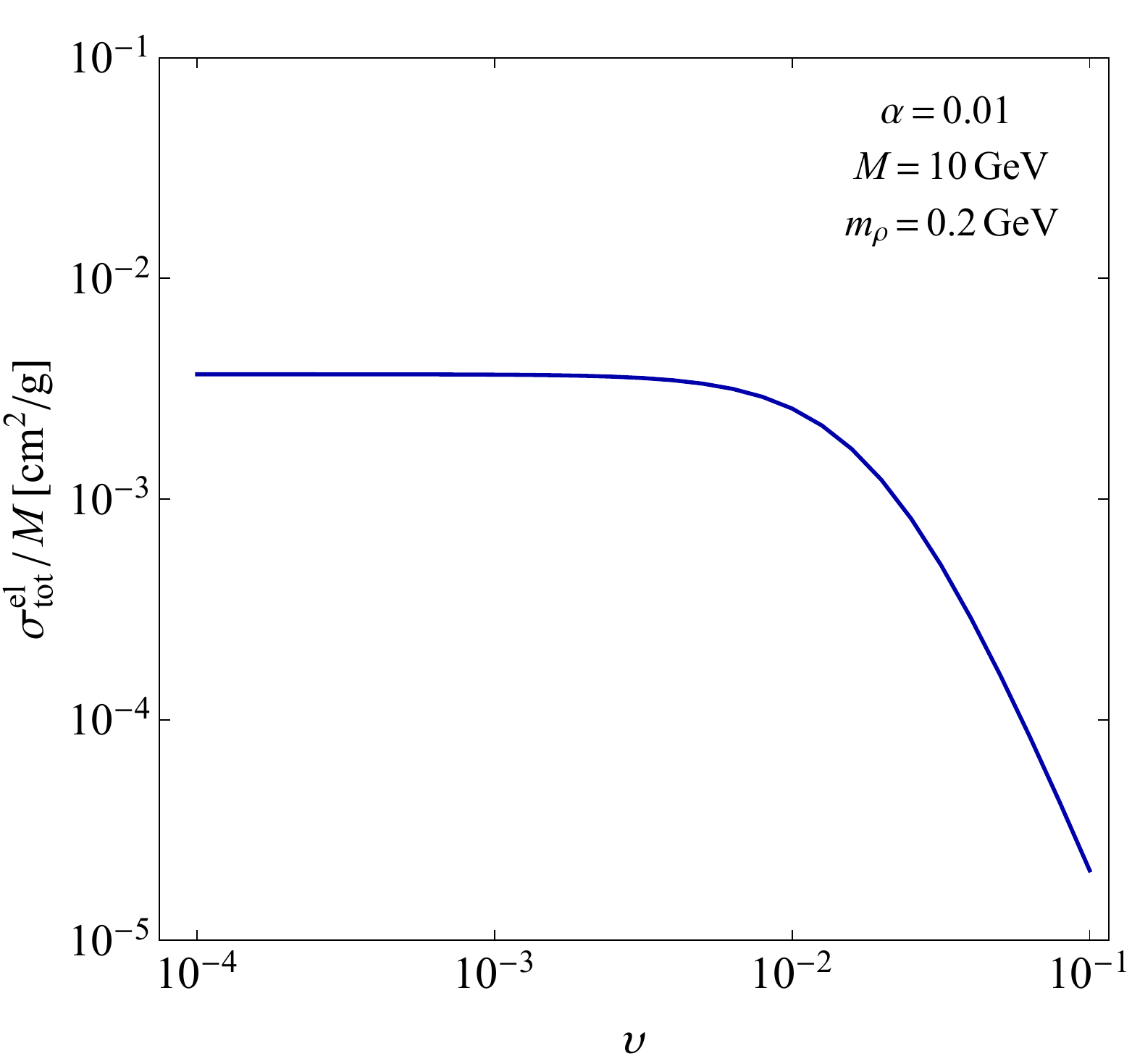}
\caption{The $\chi_1\chi_1\to\chi_1\chi_1$ cross-section in the regime $\mu_1v^2/2 < 2\Delta$, i.e., below threshold for our two-level SIDM model. This is approximately the same as in a single state SIDM model but with the potential given by Eq.(\ref{eq:diagpot}).}
\label{fig:below_thr}
\end{figure}

\begin{figure*}[!t]
 \begin{center}
  \includegraphics[width=0.4\textwidth]{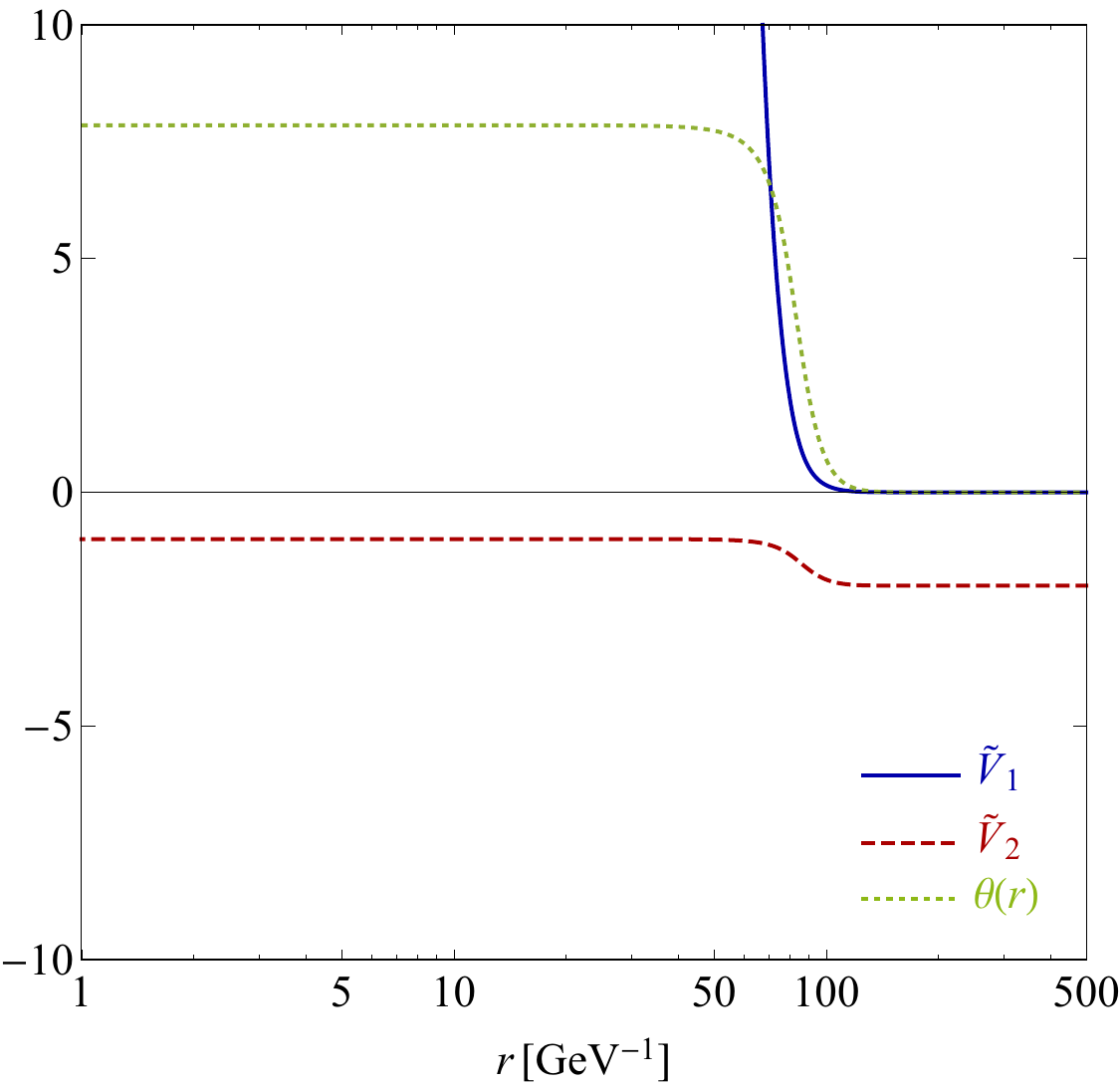}~~~~~~~~~\includegraphics[width=0.43\textwidth]{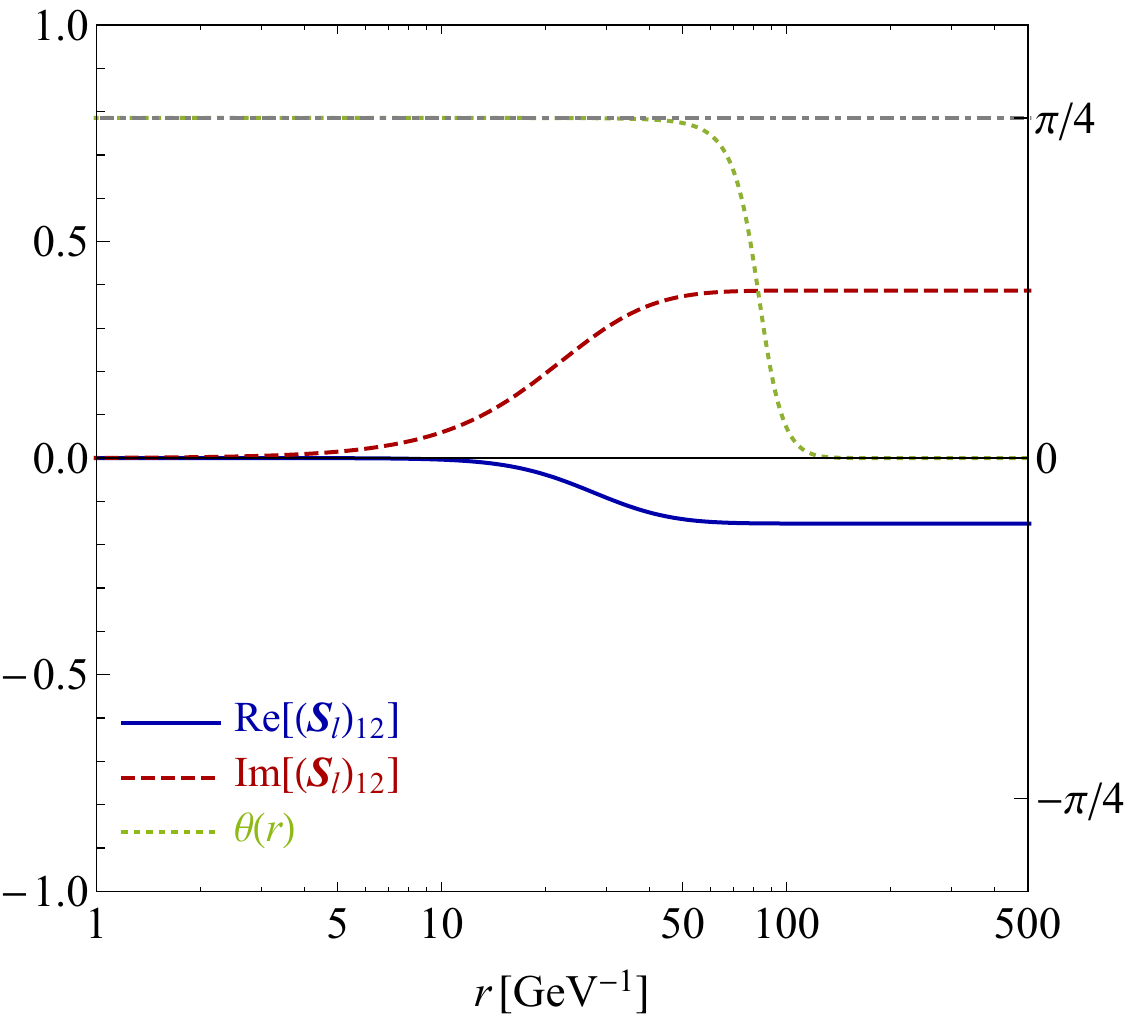}
  \caption{(Left) The eigenvalues $\tilde{V}_1$ (blue solid) and $\tilde{V}_2$ (red dashed) of the rotated potential matrix as in Eq.(\ref{eq:diagpot}). The mixing angle $\theta(r)$ is shown as the green dotted line. The unit on the vertical axis is arbitrary. Note that the eigenvalues remain well-separated and there is no level-crossing, explaining why the evolution is adiabatic. The parameters chosen here correspond to an above threshold scenario, but the behaviour of the eigenvalues does not depend on that. (Right) Real and imaginary parts of the off-diagonal component of the scattering matrix function, Re($\bS_{12}(r)$) (blue solid) and Im($\bS_{12}(r)$) (red dashed), as well as the instantaneous mixing angle $\theta(r)$ (green dotted). Note that $\theta(r)$ reaches $\pi/4$ at $r\to0$, shown as a grey dot-dashed line. The real and imaginary parts of $\bS_{12}(r)$ are shown multiplied by factors of 100 and 10, respectively, for visual clarity. Note that the off-diagonal component $\bS_{12}(r)$ deviates from its value at large $r$ only at $r\lesssim50$, where the eigenvalues are well-separated but the mixing is maximal. This shows that any nonzero inelastic scattering, that comes from a nonzero $\bS_{12}(r)$, is a result of maximal mixing and not level-mixing or lack of adiabaticity. The velocity was chosen to be 100 km/s (above threshold).}
  \label{fig:sfunc}
 \end{center}
\end{figure*}

If the total energy of the incoming particles is below the threshold, i.e., $\mu_1v^2/2 < 2\Delta$, then the excited state cannot be produced onshell as a final state, and only the elastic channel is relevant. 
The elastic cross-section shown in Fig.\,\ref{fig:below_thr} can be understood based on a simple analytical argument. In the regime $\mu_1v^2<2\Delta$, transitions to the heavier excited state are classically forbidden by the vanishing tunnelling probability and can be neglected.  As a result, it is easiest to diagonalize Eq.(\ref{eq:schro}) locally at at each value of $r$, and solve only for the elastic scattering cross-section in the ground state with the potential $\tilde{V}_1(r)$ given by
\begin{equation}\label{eq:diagpot}
\tilde{V}_{1,2} = -V_1-\Delta\pm\sqrt{V^2_1+\Delta^2}\,.
\end{equation}
Since the potentials are functions of the radial distance, the rotation angle will also be a function of $r$,
\begin{equation}
\label{eq:theta}
 \tan2\theta(r) = -V_1(r)/\Delta\,.
\end{equation}
If the spatial derivatives of $\theta(r)$ are small enough, then one can assume the system to be adiabatic\,\cite{Zhang:2016dck}, i.e., the two instantaneous energy levels do not mix.
Note that $\theta(r)\to \pi/4$ for $r\to0$ and $\theta(r)\to0$ for $r\to\infty$. Therefore the two states are completely unmixed at large distance but get maximally mixed as the incoming particles get closer. The radial dependence of $\theta(r)$ is shown in the green curve in the left panel of Fig.\,\ref{fig:sfunc}. The transition from zero to $\pi/4$ happens in the region where $|V_1(r)|\simeq\Delta$. Beyond this point towards large $r$, mixing ceases as the potential in the off-diagonal position becomes smaller than the diagonal mass gap term. This behaviour of the mixing angle allows us to use the rotated basis to determine the elastic scattering cross-section in the below threshold regime. The smallness of the radial dependence of the elements of the rotation matrix ensures that the system remains in one of the eigenstates during the complete scattering. In Fig.\,\ref{fig:sfunc}, we also show the behaviour of the eigenvalues of the potential matrix with $r$ for $\ell = 0$. The eigenvalues never cross each other and their separation goes as $\sim 2/r$ for $r \to 0$ reaching a constant $2\Delta$ for large $r$. This remains true for all higher partial waves and signals the fact that the evolution is always adiabatic and the elastic cross section is approximately given by the effective potential in Eq.(\ref{eq:diagpot}).

\subsection{Above Threshold}
If the energy of the incoming particles is sufficiently large, i.e., $\mu_1v^2/2 > 2\Delta$, then the excited state can be produced onshell. In this case, the DM particles in the ground state can upscatter inelastically to the excited state. A measure of the inelasticity in the system is given by the magnitude of the off-diagonal elements of the scattering matrix $\Sl$. We study the behaviour of the scattering matrix using the \textit{variable phase method} following Ref.\,\cite{calogero1967variable}. In this method the wavefunction is first written in an integral form,
\begin{widetext}
\begin{equation}\label{eq:integral}
 \Psil(r)=\left(\frac{\mu}{k}\right)^{1/2}\Jl(kr)-\frac{2\mu}{k}\int_0^r dt \left(\Jl(kr)\Nl(kt)-\Jl(kt)\Nl(kr)\right)V(t)\Psil(t)
\end{equation}
\end{widetext}
with the Riccati-Bessel functions $\Jl(kr)$ and $\Nl(kr)$ as defined in Eq.(\ref{eq:bessel}). In order to isolate the part of the wavefunction arising from the interaction potential $V(r)$, it helps to define another function $\Fl(r)$ as
\begin{equation}\label{eq:F}
 \Fl(r)=\frac{1}{2}\left[\mathit{1}+2\int^r_0 dt\,\Hl^{(2)}(t)V(t)\Psil(t)\right]\,.
\end{equation}
On substitution of this in Eq.(\ref{eq:integral}), one gets
\begin{equation}
 \Psil(r)=-i\left[\Hl^{(1)}(r)\Fl(r)-\Hl^{(2)}(r)\Fl^*(r)\right]\,,
\end{equation}
where one can clearly identify $\Hl^{(1),(2)}\equiv(\mu/k)^{1/2}(\Nl\pm i\Jl)$ as the free wave solutions and the unknown functions $\Fl$ and $\Fl^*$ as the \textit{distortions} to the plane-wave solution due to the potential. The scattering matrix function $\bS_\ell(r)$ is defined in terms of $\Fl(r)$ as
\begin{equation}
 \bS_\ell(r)\equiv \Fl(r)\Fl^*(r)^{-1}\,.
 \label{eq:FlFl}
\end{equation}

The significance of the function $\bS_\ell(r)$ lies in the fact that its asymptotic value at large $r$ yields the scattering matrix $\Sl$. The differential equation for $\bS_\ell$ is easily obtained, by taking a derivative of the previous equation and using Eq.(\ref{eq:integral}, \ref{eq:F}),
\begin{equation}
 \frac{d\bS_\ell}{dr} = i\left(\bS_\ell\Hl^{(1)}-\Hl^{(2)}\right)V\left(\Hl^{(1)}\bS_\ell-\Hl^{(2)}\right)\,.
\end{equation}
The initial condition is $\bS_\ell(0)=\mathit{1}$, because at $r\to0$ the function $\bS_\ell(r)$, as given by Eqs.\,(\ref{eq:FlFl}) and (\ref{eq:F}), has zero off-diagonal entries because the integral in Eq.(\ref{eq:F}) vanishes.

The off-diagonal components of $\bS_\ell(r)$ track the behaviour of the inelastic cross-section with $r$. In the right panel of Fig.\,\ref{fig:sfunc}, we plot the real and the imaginary parts of the off-diagonal elements of $\bS_\ell$ as a function of $r$, along with the rotation angle $\theta(r)$ of the potential matrix $V(r)$ in Eq.(\ref{eq:theta}). The inelastic cross-section grows from zero to its asymptotic value during the region where the mixing angle is $\pi/4$ (maximal mixing). More precisely, it saturates at around $r \sim 1/m_\rho$ (In this case, $m_\rho = 0.1$ GeV). Beyond this point the off-diagonal potentials are exponentially screened and the off-diagonal coupling between two states vanishes. Also note that nothing special happens in the `nonadiabatic' region, i.e., where the angle varies from $\pi/4$ to zero towards large $r$. While we show this for particular values of the parameters in the potential, this qualitative behaviour does not change for other values of the parameters. Therefore, one can conclude that the inelasticity is driven by the maximal mixing between two states near the origin (the adiabatic mixing) which yields the large upscattering cross-section from the ground state, and not by the nonadiabaticity in the system. As soon as this mixing goes to zero, the inelastic cross-section saturates to its asymptotic value. We also note that the asymptotic value of the off-diagonal elements of the $\bS_\ell$ matrix is significantly large, which hints towards a large inelastic cross-section in the presence of an off-diagonal potential. For obtaining our numerical results, we have used the method shown in Appendix\,\ref{sec:appendix}, but the main advantage of the variable phase method described here is that it reveals that the origin of inelasticity is large mixing, not non-adiabaticity. A variant of this method was previously used to compute the cross-sections in Ref.\,\cite{Blennow:2016gde}.
\begin{figure*}[!t]
\begin{center}
 \includegraphics[width=0.4\textwidth]{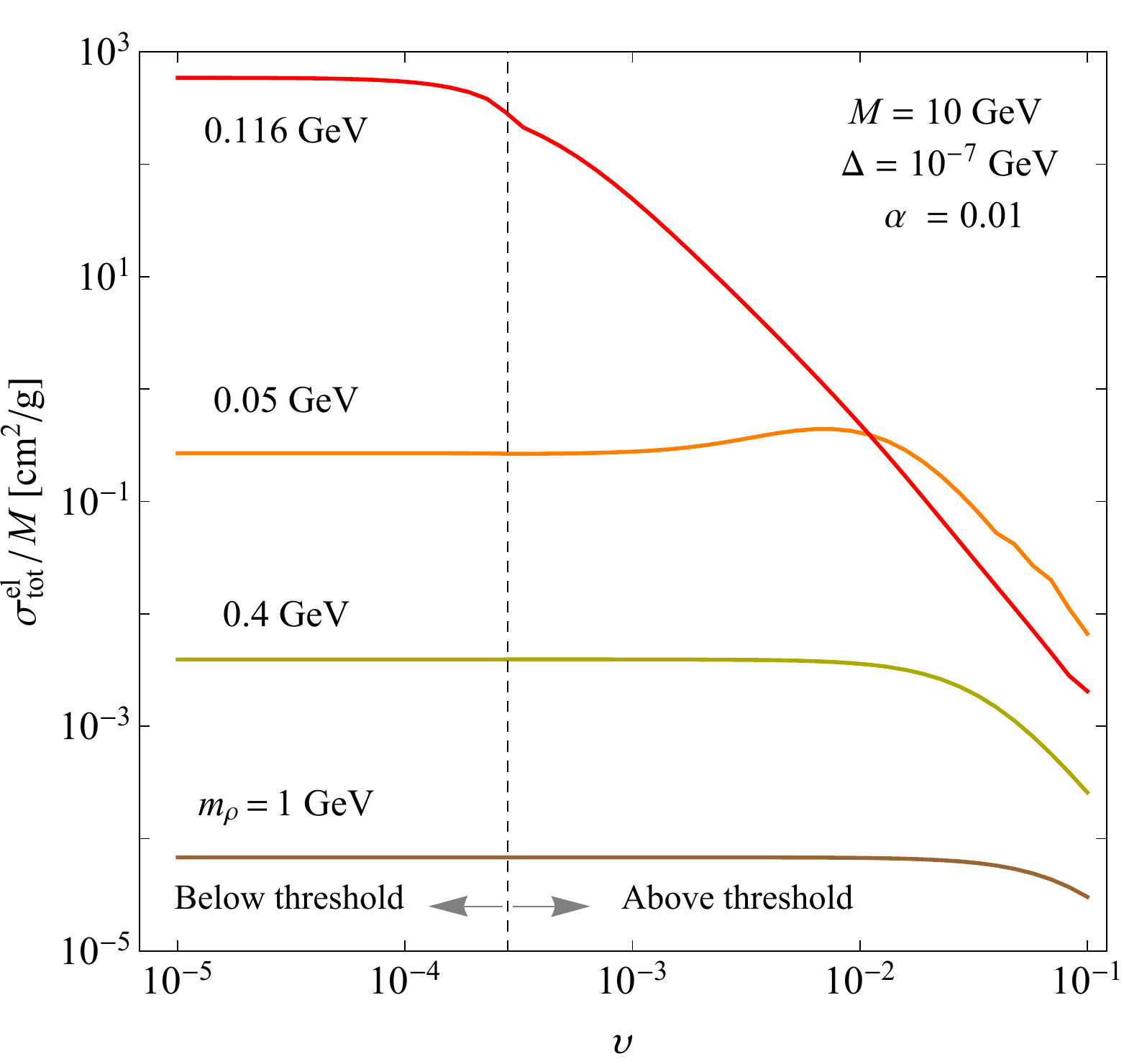}~~~~~\includegraphics[width=0.4\textwidth]{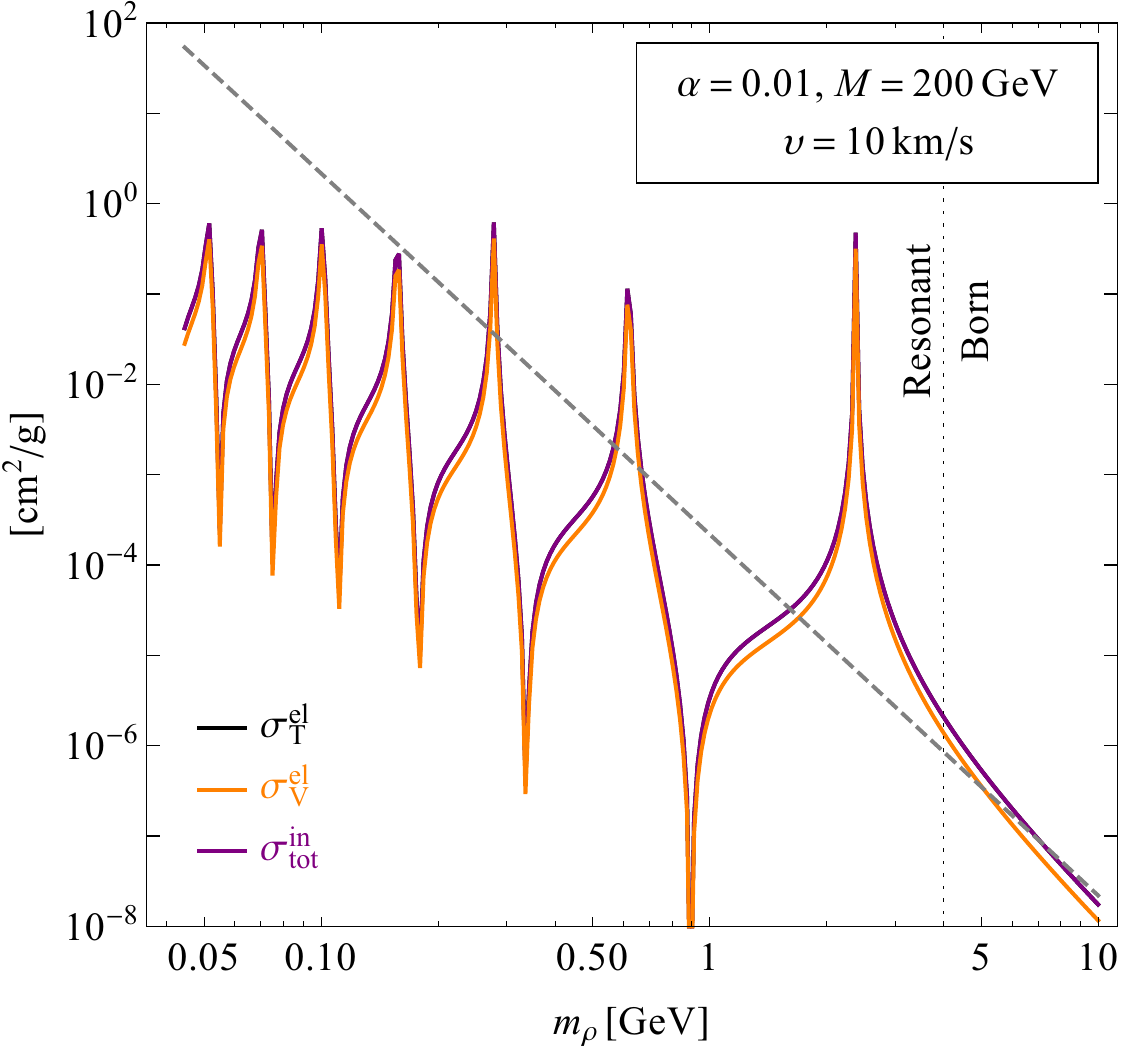}
 \caption{(Left) The velocity dependence of the elastic self-scattering cross-section for different values of $m_\rho$ as indicated in the figure. (Right) The elastic transfer and viscosity cross-sections for a particular choice of the parameter values: $M = 200{\rm\ GeV}, \alpha = 0.01, v = 10{\rm\ km/s}$. The grey dashed line shows the analytical estimate of the Born cross-section for the 2-level model, obtained using the 1-level equivalent proposed in Ref.\,\cite{Das:2016ced} with the substitution $\alpha\to2\alpha$ explained in text, and the numerical results are in good agreement with the analytical estimate of the Born cross section in the large $m_\rho$ limit. The resonant values of $m_\rho$, given in Eq.(\ref{eq:resonance}), similarly agree very well.}
 \label{fig:elin}
\end{center}
\end{figure*}

The left panel of Fig.\,\ref{fig:elin} shows the velocity-dependence of the elastic cross-section for a few representative values of the mediator mass $m_\rho$, as indicated in the figure. In general, we see that irrespective of the value of $m_\rho$, the $\sigma_{\rm el}$ is larger for small DM velocity and decreases for large velocity. Therefore it is possible to enhance the self-scattering in the dwarf-sized objects and address the core-cusp problem, while simultaneously suppressing it in the larger cluster-size objects and satisfying the upper bounds coming from colliding clusters\,\cite{Dawson:2011kf, Kim:2016ujt}. The values of $m_\rho$ were chosen such that they span across a resonance. The curve labelled by $m_\rho = 0.116$ GeV corresponds to a resonance in the cross-section and hence shows large enhancement in the small velocity regime. On either side of this resonance, the cross-section decreases. These features are unaltered relative to single-level DM models.

In the right panel of Fig.\,\ref{fig:elin}, we show the behaviour of the elastic transfer and viscosity cross-sections with the mediator mass in the $\Delta \to 0$ limit. 
Two distinct regions, Born ($\alpha M/m_\rho \ll 1$) and resonant ($\alpha M/m_\rho \gtrsim 1$), are apparent. The dashed grey line shows an approximate analytical estimate of the cross-section in the Born limit \cite{Feng:2009hw}. Although the physical system in this work is different than that in single-state DM models, it is possible to get an approximate expression of the Born cross-section by a substitution $\alpha \to 2\alpha$. This substitution is based on the understanding that in the limit of small $\Delta$, the two states $|\chi_1\chi_1\rangle$ and $|\chi_2\chi_2\rangle$ become indistinguishable and related to each other through the relation $|\chi_2\chi_2\rangle=(-1)^{\ell+s}|\chi_1\chi_1\rangle$ where $s$ is the total spin of the state, as was explained in a previous paper by us\,\cite{Das:2016ced}. Only one linear combination of the states survives in this limit and one can map the two-level system onto a single state with an effective potential
\begin{equation}
 V_{\rm eff} = V_{11} + (-1)^{\ell+s}V_{12}\,.
\end{equation}
This also explains the substitution $\alpha\to 2\alpha$ in our adapted estimate of the resonant condition given by \cite{Cassel:2009wt}, 
\begin{equation}
\label{eq:resonance}
 \frac{2\alpha M}{1.64\, m_\rho} = n^2, \qquad n = 1, 2, 3\cdots\,,
\end{equation}
that explains the positions of the resonances and agrees very well with the numerical results. 

\begin{figure}[!b]
 \begin{center}
  \includegraphics[width=0.9\columnwidth]{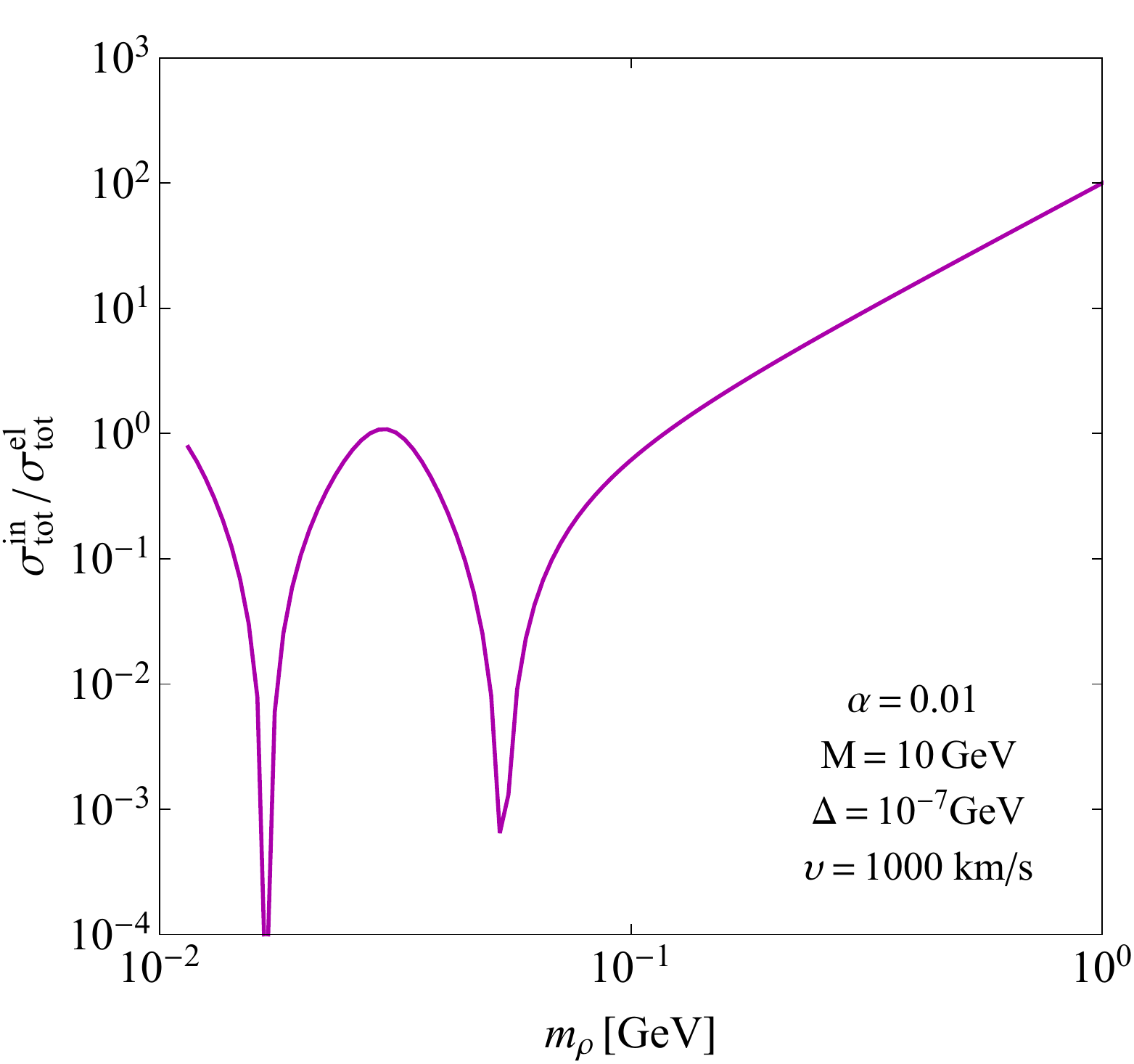}
  \caption{The ratio of the inelastic to the elastic cross-sections in the purely off-diagonal interaction DM model, with parameters as indicated in the figure, showing that the inelastic cross-section can be much smaller or much larger than the elastic cross-section in strongly off-diagonal models.}
	\label{fig:ratio}
 \end{center}
\end{figure}

In all these cases the inelastic cross-section is almost equal to the elastic cross-section as shown in the right panel of Fig.\,\ref{fig:elin}. This strong correlation between the two cross-sections is a result of setting all components of the potential matrix in Eq.(\ref{eq:potential}) to be equal. If the diagonal potentials are weaker than the off-diagonal counterparts, then the two cross-sections can be different by several orders of magnitude. An extreme example of such a case is the model of a two-component Majorana DM particles charged under a broken U(1) gauge symmetry\,\cite{Schutz:2014nka, Blennow:2016gde}. The conserved currents in this model are given by $\bar{\chi}_1\gamma^\mu\chi_2$ and $\bar{\chi}_2\gamma^\mu\chi_1$. Hence at tree level, elastic scattering $\chi_{1,2}\chi_{1,2} \to \chi_{1,2}\chi_{1,2}$ is not possible although inelastic scattering $\chi_{1,2}\chi_{1,2} \to \chi_{2,1}\chi_{2,1}$ can take place through an exchange of a gauge particle. The lowest order elastic process will involve at least one loop. Therefore in the Born limit (large $m_\rho$), the elastic cross-section will be suppressed. In the resonant and classical regime these interactions would give rise to purely off-diagonal attractive Yukawa potentials and both elastic and inelastic cross-sections will be comparable (see Fig.\,\ref{fig:ratio}). However, depending on the parameter values, one may be larger than the other.



\section{\label{sec:result}Signatures of Inelastic Scattering}

The new effects of this dissipation mechanism is driven by the large inelastic scattering rate given by $\Gamma_{\rm up}\equiv n_\chi\sigma_{\rm in}v$. An order of magnitude estimate of the timescale associated with the upscattering rate is given by
\begin{equation}
\label{eq:tup}
\begin{array}{rcl}
 t_{\rm up}&\simeq& 10^{12}{\rm\ yrs}\ \dfrac{10^4 {\rm\ M_\odot kpc^{-3}}}{\rho}\,\dfrac{\rm 1\ cm^2 g^{-1}}{\sigma_{\rm in}/M}\,\dfrac{10^3{\rm\ km s^{-1}}}{v}\,.
\end{array}
\end{equation}
The DM velocity was chosen to be $\mathcal{O}(1000)$ km/s so that the upscattering and decay processes are kinematically allowed. Clearly this typical timescale is 1-2 orders of magnitude larger than the age of the Universe, whereas large DM densities required for upscattering to take place have only been present for a much shorter time (only since nonlinear structures have formed). Therefore the effects of these upscatterings cannot be too large. We now discuss two possible effects due to this inelastic scattering.

\subsection{\label{sec:result1}Halo Cooling}
If the upscattering rate $n_\chi\sigma_{\rm in}v$ is not too small, $\chi_1$ can upscatter to the excited state $\chi_2$  and thus produced $\chi_2$ will promptly decay into the light mediator particle $\rho$ and $\chi_1$. If the $\chi_1-\rho$ scattering cross-section is small in the given DM halo, then these light particles may escape the halo, thereby cooling the halo at a considerable rate. Large upscattering requires the colliding DM particles to be energetic enough so that sufficient phase space is available for the excited state. For example, a 10 GeV mass DM with $\Delta =$ 1 MeV has a velocity threshold of $\sim$ 1000 km/s. Thus, this phenomenon will mostly be important in objects with large DM velocity dispersions, e.g., in large galaxies and galaxy clusters.

A thorough analysis of the effect of this dissipation mechanism on the DM halo structure does not yet exist in the literature. We will not attempt a full treatment here. However, a qualitative understanding can be gained from the response of DM halos for similar cooling processes present in the baryonic matter, as we recap below. After falling towards the centre of a halo, the baryons interact with each other and condense into lower energy states. In the process, the particles dissipate away a considerable amount of energy in the form of radiation which may escape the halo. The less energetic baryons then condense and undergo further infall towards the centre. The changing shape of the baryon density profile affects the DM profile by increasing density near the centre. The analytical estimations of this effect have been worked out in the adiabatic contraction regime\,\cite{Blumenthal:1985qy}. In this regime, the DM particle orbits are assumed to be circular or nearly circular and the total mass enclosed by the orbit is assumed to be changing very slowly compared to the orbital time period of the DM particle. In this adiabatic regime, the invariance of $\oint pdq$ implies
\begin{equation}
 M(r)r = {\rm constant}\,.
\end{equation}
Here $M(r)$ is the total mass enclosed inside the orbit of radius $r$. Using this invariance, an analytical estimate has been obtained which fairly matches with the numerical N-body simulation results\,\cite{Blumenthal:1985qy, 2004ApJ...616...16G}. The main effect is the steepening of the DM density profile near the centre and forming a denser core. As more baryons fall towards the centre, the gravitational well becomes deeper and more DM particles are attracted inward. This increases the slope of the central density profile\,\cite{2004ApJ...616...16G}.

In our case, the DM component itself will have a dissipation or cooling mechanism through an upscattering and a subsequent decay of the excited state. This process is independent of and in addition to the baryonic cooling. Hence the effect of halo cooling will presumably be more prominent in this scenario and one would expect more complexity and richness in the small-scale structure of DM. As a result larger portion of the parameter space can be constrained. 

The rate of this new dissipation mechanism will mainly be given by the upscattering rate as the the decay is very fast and can be assumed to be prompt. Here we will give a rough estimate of the rate of energy loss due to the upscattering and decay from the excited state. In the limit of nonrelativistic DM  and $\Delta \ll M$, the net kinetic energy lost per particle is approximately equal to $\Delta$ itself. The upscattered $\chi_2$ particles will decay and produce lighter particles with some amount of kinetic energy from the phase space available. One can estimate the leading order contribution to this energy gain to be $\mathcal{O}(\Delta^2/M^2)$ and $\mathcal{O}(v^2_2\Delta/M)$ where $v_2$ is velocity of the upscattered $\chi_2$ particles prior to decay. Therefore for all relevant parameter choices, the gain in the kinetic energy from the decay is negligible relative to the energy loss from the upscattering. The requirement for the upscattering and the decay to happen constrains the parameter space as $\mu_1v^2/4 > \Delta > m_\rho$. We shall assume that all light particles generated from the decays leave the halo.

In a halo, the average rate of energy loss in a DM shell of radius $r$ and width $dr$, is estimated by
\begin{eqnarray}
 &4\pi r^2dr\,\Gamma_{\rm up}(r) n_\chi(r)\, 2\Delta 
 =& 4\pi r^2dr\,\frac{2\Delta}{M}\frac{\sigma_{\rm in}}{M}\,v\rho(r)^2\,.
\end{eqnarray}
The radial dependence of DM velocity could be estimated from simple Newtonian dynamics. It peaks around the scale radius of the halo with an NFW density profile. The NFW profile is defined as follows
\begin{equation}
 \rho_{\rm NFW}=\frac{\rho_s}{\dfrac{r}{r_s}\left(1+\dfrac{r}{r_s}\right)^2}
\end{equation}
where $\rho_s$ and $r_s$ are the scale density and radius, respectively.
The individual DM velocities at a given radius will follow a thermalized Maxwell-Boltzmann (MB) distribution characterized by a virial velocity dispersion $\bar{v}(r)$. Essentially, in a fully virialized halo, the high energetic DM particles will most often occupy the outer edges of the halo. The halo cooling rate will be given by a convolution over the DM velocity distribution function
\begin{equation}\label{eq:cooling}
\begin{array}{rcl}
 \dfrac{dE}{dt} &=& 4\pi r^2 dr\,\dfrac{2\Delta}{M}\rho(r)^2\displaystyle\int^\infty_0 \dfrac{\sigma_{\rm in}}{M}\bar{v}(r)f(v)dv\,,
\end{array}
\end{equation}
where we take $f(v)$ to be approximated by a Maxwell distribution $f_{\rm MB}(v) = 4\pi v^2 \exp\left[-v^2/\bar{v}(r)^2\right]/(\sqrt{\pi}\bar{v}(r))^3$. Note that the velocity distribution $f(v)$ also depends on radial distance $r$ through $\bar{v}(r)$.

\begin{figure}[!t]
 \begin{center}
 \vspace{0.2cm}
  \includegraphics[width=0.86\columnwidth]{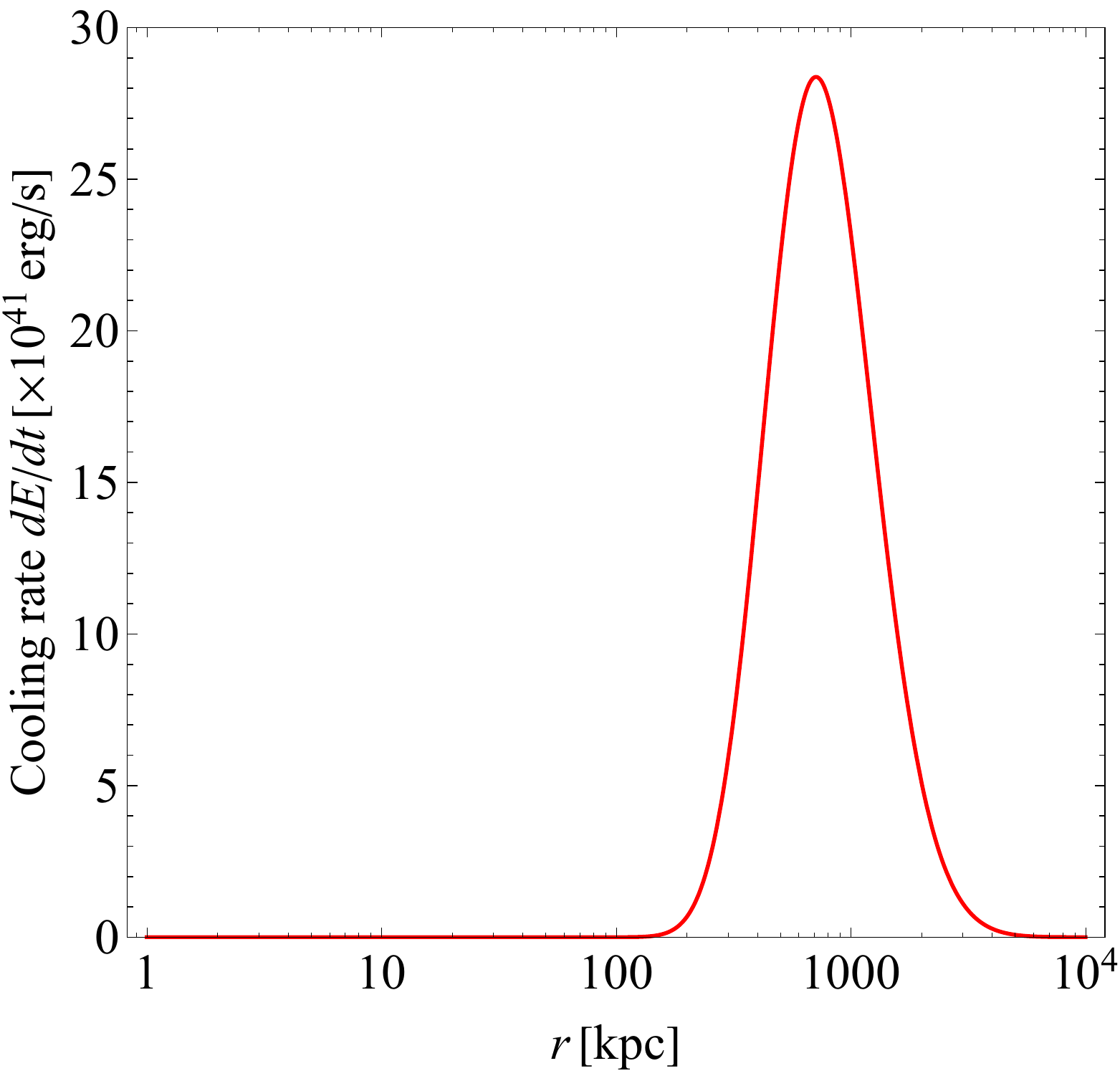}
  \caption{The radial dependence of the cooling rate $dE/dt$ (see Eq.(\ref{eq:cooling})) of a Virgo cluster-size halo with a scale radius $r_s=560$ kpc and density $\rho_s=3.2\times10^5\, {\rm M_\odot/kpc^3}$. The chosen DM parameters are $M=10$ GeV, $\Delta= 10^{-4}$ GeV and $\sigma_{\rm in}/M=1\ {\rm cm^2/g}$.}
  \label{fig:cooling}
 \end{center}
\end{figure}

An approximate radial dependence of the cooling rate $dE/dt$ for a halo of the size of that of the Virgo cluster is shown in Fig.\,\ref{fig:cooling}. The profile was taken to be an NFW with a scale radius $r_s = 560$ kpc and density $\rho_s = 3.2 \times 10^5\ {\rm M_\odot/kpc^3}$. For simplicity the inelastic cross-section was taken to be velocity-independent constant $\sigma_{\rm in}/M = 1\ {\rm cm^2/g}$. The resulting cooling rate shows a strong radial dependence and is largest near the virial radius. This cooling rate is to be compared with the energy inflow from the gravo-thermal collapse of the DM particles and due to the heat diffusion through self-scattering. The gravitational collapse brings faster (hotter) particles from the outer region of the halo to cooler inner part. And the scattering between the particles help diffuse the kinetic energy from the hotter particles to the relatively colder ones. The process of gravo-thermal collapse can be modelled following the Refs.\,\cite{1980MNRAS.191..483L, Pollack:2014rja}. The negative specific heat of a halo after virialization leads to this collapse. If we treat the DM particles as a fluid, the heat radiated inward at some radius $r$ is given by
\begin{equation}
\label{eq:inflow}
 \frac{L}{4\pi r^2} = -\frac{3}{2}abv\sigma\left(a\sigma^2+\frac{b}{C}\frac{4\pi G}{\rho v^2}\right)^{-1}\frac{\partial v^2}{\partial r}\,.
\end{equation}
Here the two terms within parenthesis on the RHS correspond to two different \textit{mean free path} regimes. The first term describes the \textit{hard sphere scattering} with the dimensionless coefficient $a = \sqrt{16/\pi}$. The second term describes the \textit{short mean free path} regime which is proportional to the gravitational constant and the numbers $b = 25\sqrt{\pi}/32$ and $C \approx 0.75$. Typical values of this heat inflow rate are 2-3 orders of magnitude larger than the cooling rate discussed above. Nevertheless, in the resonant region of the parameter space this halo cooling is expected to be efficient enough to distort the halo.

Upscattering and decay do not start abruptly but rather are continuous processes which will be present during the virialization process of the halo. At the initial epoch of structure formation the DM particles are highly nonrelativistic and there will be no dissipation. After DM falls towards the centres of the potential wells, acquires more energy and inelastic collisions become possible, it leads to cooling. From Eq.(\ref{eq:tup}) it is clear that the inelastic scattering is a rather slow process and the halo will virialize at a faster rate than the dissipation. As a result, subsequent changes in the halo shapes is expected to be continuous and not episodic. A more detailed study of the effect of this new cooling mechanism will require an N-body simulation with this extra energy loss implemented in the dark sector\,\cite{Dubinski:1993df}.

A similar halo cooling mechanism was considered in Ref.\,\cite{Boddy:2016bbu} in the context of an atomic DM model. There neutral atomic dark hydrogen makes the DM abundance in the present Universe. The hyperfine splitting in the ground state of the dark atom leads to inelasticity in the system and the excited state decays to the ground state emitting massless dark photons. The masslessness of the dark photon implies that this cooling mechanism is more important for smaller halos because of their lower gravitational binding energy. On the contrary, in our case the particle $\rho$ is massive. Hence the cooling mechanism shuts off for small mass halos where the DM particles do not have enough energy to upscatter, and the dissipation arises mainly in large galaxies or clusters. Note that the details of the particle physics model do not affect the radial dependence shown in Fig.\,\ref{fig:cooling}, and all such details are encapsulated into the velocity dependence of the cross-section that determines this feature.

\subsection{\label{sec:result2}Drag and Evaporation from Inelastic Scattering}
Independent bounds on DM scattering can be obtained from particle evaporation during collision of clusters and the movement of smaller dwarf-size halos through larger halos\,\cite{Kahlhoefer:2013dca, Kummer:2017bhr}. 
As was pointed out in Ref.\,\cite{Kahlhoefer:2013dca}, the SIDM particles will experience collisions in colliding clusters, whereas the stellar components of the objects will move freely without any appreciable friction. If the momentum transfer in a DM-DM collision is such that the final velocity is larger than the escape velocity of the parent halo then it would leave the halo and would result in DM \textit{evaporation} from the halo. The existing observations from colliding clusters put strong constraint on this process yielding an upper bound on the DM self-scattering cross-section. An estimate of the of rate such collisions can be obtained following Ref.\,\cite{Kahlhoefer:2013dca}, in the limit of long-range interaction (as the hierarchy $\mu_1v^2/4 > \Delta > m_\rho$ is easy to satisfy with smaller value of $m_\rho$ even at cluster size scale). In Ref.\,\cite{Kahlhoefer:2013dca}, the cumulative evaporation rate was shown to be more important than the immediate evaporation when DM has long range self-interaction. This rate is given by
\begin{equation}
 \label{eq:cmlrate}
 R_{\rm cml} = \frac{\eta \alpha^2\rho_{\rm DM}}{M^3 v^3_0}\left[1-2\log\left(\frac{\theta_{\rm min}}{2}\right)\right]\,.
\end{equation}
Here $v_0$ is the relative velocity between the two colliding clusters and $\rho_{\rm DM}$ is the DM density in the bigger halo. The parameter $\theta_{\rm min}$ encodes the screening length and regulates the forward divergence. Because of this evaporation rate, the clusters will feel a drag force given by
\\
\begin{align}
 \label{eq:drag1}
 \frac{F_{\rm drag}}{M} &= v_0R_{\rm cml}\\
  &= \frac{\eta \alpha^2\rho_{\rm DM}}{M^3 v^2_0}\left[1-2\log\left(\frac{\theta_{\rm min}}{2}\right)\right] = \frac{\tilde{\sigma}\rho_{\rm DM}}{4M v^2_0}\,,
\end{align}
where $\eta$ is an $\mathcal{O}(1)$ numerical factor depending on the nature of the mediator. In the last equality, following Ref.\,\cite{Kahlhoefer:2013dca}, we have defined the cross-section $\tilde{\sigma}$ as
\begin{equation}
 \frac{\tilde{\sigma}}{M} \equiv \frac{4\eta\alpha^2}{M^3}\left[1-2\log\left(\frac{\theta_{\rm min}}{2}\right)\right]\,.
\end{equation}
The existing bound on $\tilde{\sigma}$ from the abundance of dwarfs in our MW halo is very strong, $\tilde{\sigma}/M \lesssim 10^{-11}\ {\rm cm^2/g}$ \cite{Kahlhoefer:2013dca}.

For two-level DM, two distinct cases may arise. Firstly the usual evaporation of DM particles is still feasible in this model, and has contributions from both elastic and inelastic scatterings. If the velocities of the scattered particles are larger than the escape velocities then they can escape the halo and would collectively cause dynamical friction between the halos. Second, inelastic scattering and subsequent decay provides an additional way for energy dissipation and gives an additional contribution to the drag force. For simplicity, if we assume that all DM particles are moving at the same velocity $v_0$ as the halo then $n_\chi\sigma_{\rm in}v_0$ is the upscattering rate per unit time. After each upscattering and decay event, two light particles escape the halo taking away an amount of energy which is roughly $\langle E_{\rm decay}\rangle \simeq 2\Delta$. Therefore, the halo loses energy at a rate $dE/dt$,
\begin{equation}
 \frac{dE}{dt} = \langle E_{\rm decay}\rangle n_\chi\sigma_{\rm in}v_0\,.
\end{equation}
The resulting drag force per unit DM mass (or deceleration) due to this energy loss is given by
\begin{equation}
 \frac{F^{\rm decay}_{\rm drag}}{M} = \frac{1}{Mv_0}\frac{dE}{dt} = \frac{\langle E_{\rm decay}\rangle}{M}\frac{\rho_{\rm DM}\sigma_{\rm in}}{M}\,.
\end{equation}
Then the net drag force acting between the halos is given by
\begin{align}
 \label{eq:drag2}
 \dfrac{F_{\rm drag}}{M} &= v_0R_{\rm cml} + \dfrac{\langle E_{\rm decay}\rangle}{M}\dfrac{\rho_{\rm DM}\sigma_{\rm in}}{M}\\
 &=\dfrac{(\tilde{\sigma}_{\rm el}+\tilde{\sigma}_{\rm in})\rho_{\rm DM}}{4M v^2_0} + \dfrac{2\Delta}{M}\dfrac{\rho_{\rm DM}\sigma_{\rm in}}{M}\,.
\end{align}
The first term on the r.h.s above represents the cumulative evaporation rate, due to elastic and inelastic processes that are approximately equal across a large portion of the parameter space. We neglect the tiny velocity gain of $\chi_1$ from the decay as we have seen it to be of even smaller order of magnitude in the previous subsection. The second term corresponds to the new dissipation mechanism from upscattering and decay. The quantity $\langle E_{\rm decay}\rangle$ denotes the energy loss rate averaged over the phase space of the final particles which, in the last equality, has been approximated to $\langle E_{\rm decay}\rangle\simeq 2\Delta$. For simplicity here we have assumed that all DM particles in the incident halo have velocity $v_0$. Of course a more careful analysis would require averaging over a Maxwellian distribution characterized by a velocity dispersion $v_0$.

The relative size of the new term in Eq.(\ref{eq:drag2}) compared to the old term is given by $\sim 4 v^2_0\Delta/M \simeq 10^{-8}$ for $M = 10$ GeV, $\Delta = 1$ MeV and $v_0 = 1000$ km/s and assuming $\sigma_{\rm in} \simeq \tilde{\sigma}_{\rm in}$. The parametric smallness of the new drag force term may be traced back to the smallness of the mediator mass. A light particle-mediated interaction has a negative power dependence on velocity and is enhanced at small velocities, whereas the new term is virtually velocity independent. This velocity dependence may be useful to extract the impact of the second term, relative to the larger first term. We leave this investigation to a more detailed study.

There may be other signatures of this energy loss process. For example, just as the baryonic energy loss processes like Compton scattering and bremsstrahlung are responsible for the collapse of the ordinary matter into disk-like structures forming the galaxies, for two-level DM, upscattering and subsequent decay processes help DM lose energy and can lead to the formation of a rotating {dark disk} in DM halo\,\cite{Read:2008fh, Purcell:2009yp, Cholis:2010px, Fan:2013yva, Fan:2013tia, McCullough:2013jma}. As another signature, the authors in Ref.\,\cite{More:2016vgs}, observed a discrepancy between the predicted positions of the \textit{splashback radii} (see\,\cite{Fillmore:1984wk, Bertschinger:1985pd, Adhikari:2014lna}) of cluster-size halos in simulation and the observational data\,\cite{redmapper, 2014ApJ...785..104R}. This mismatch could in principle be addressed by this energy dissipation mechanism through DM inelastic scattering. 


\section{\label{sec:concl}Summary \& Outlook}
In this work, we have studied the self-scattering of a two-level DM model. The off-diagonal interaction leads to inelastic scattering of a pair of DM particles from the ground state to the excited state, in addition to the ordinary elastic scattering.

If the incoming energy of the particles is below threshold, the excited state is not produced as physical states. Nevertheless, those states are produced offshell in the intermediate steps of the scattering and can affect even the elastic scattering cross-section. It was shown that the equations in this case, can be rotated to a new basis where the potential matrix becomes diagonal, and because of adiabaticity can be solved as a single state system with an appropriate potential.

When the incoming particles are above threshold, inelastic scattering may also take place. We have shown that in a large part of the parameter space, the inelastic cross-section is comparable to its elastic counterpart. This large inelasticity is a result of the maximal adiabatic mixing between the two states. We have also identified the Born and resonant regions in the relevant parameter space, and an estimate for the resonance condition has been given using a mapping of the two-level system to an equivalent one-level equation.

The off-diagonal interaction between the DM states allows the heavier state decay to the lighter one and the mediator. The upscattering and subsequent decay thus provides a mechanism for energy dissipation in DM halos. Assuming the decay to be prompt, the rate of the upscattering induced decay is given by the inelastic scattering rate which we computed to be 1-2 orders of magnitudes larger than the age of the Universe. Therefore, the DM halos can not condense into smaller halos via this mechanism. Rather, the inelastic process takes place only in larger objects and is effective only after the DM density becomes large enough at the centres of those objects. We compared this cooling rate with the heating due to ordinary elastic scattering and found that in some regions of the parameter space, the cooling rate could be a large fraction of the heating rate. We expect that this will leave an observable imprint on DM halo formation and evolution which can be only be probed by an N-body simulation incorporating this dissipative feature.

The same dissipation gives rise to an additional drag force between two colliding halos or for a small halos drifting through a larger one. When two halos collide with each other, the self-interacting DM particles scatter with each other and lose energy by emitting the light scalars. This energy loss can be interpreted as a new drag force acting between the halos. We calculated an analytical expression for this new drag force and found that it is small relative to the other contribution from ordinary scattering, but has a distinctive velocity independence unlike the usual drag force.

\section{Acknowledgments}
We thank Susmita Adhikari, Arka Banerjee, Mattias Blennow, Amol Dighe, Bhuvnesh Jain, Felix Kahlhoefer, Surhud More, and Annika H. G. Peter for helpful discussions.  This work was partially funded through a Ramanujan Fellowship of the Dept. of Science and Technology, Government of India, and the Max-Planck-Partnergroup ``Astroparticle Physics'' of the Max-Planck-Gesellschaft awarded to B.D and has received partial support from the European Union's Horizon 2020 research and innovation programme under the Marie-Sklodowska-Curie grant agreement Nos. 674896 and 690575. The work of A.D. was supported by the Dept. of Atomic Energy, Government of India.

\appendix
\section{\label{sec:appendix}Formalism for Multi-channel Scattering}
In a general case of an $N$-level system, the inner products of all possible 2-body states can be arranged in an $N\times N$-matrix $\Psil(r)$\,\cite{PhysRevA.49.2587}. The columns of $\Psil(r)$ denote the linearly independent regular solutions of the Schr\"odinger equation
\begin{equation}
 \left[\dfrac{1}{r^2}\dfrac{d}{dr}\left(r^2\dfrac{d}{dr}\right)+k^2-\dfrac{\ell(\ell+1)}{r^2}-2\mu V(r)\right]\Psil(r) = 0\,,
\end{equation}
where $k$ and $\mu$ are two diagonal matrices with channel momenta and reduced masses as defined in Eq.(\ref{eq:momentum}). These set of equations are supplemented by the boundary conditions at $r=r_0$ as follows,
\begin{equation}
\label{eq:bc}
 [\Psil(r_0)]_{ab}=r_0\,\delta_{ab},\qquad [\Psil'(r_0)]_{ab}=(\ell+1)\,\delta_{ab}\,.
\end{equation}
The initial point $r_0$ is chosen to be small enough so that the centrifugal term dominates over the other two terms in the differential equation. The overall normalization is irrelevant as we are interested only in the final cross-section. Numerically, we start solving the equations at $r = r_0$ and proceed towards larger $r$. We choose a sufficiently large $r =r_f$ where the potential becomes negligible compared to the kinetic energy term. At $r=r_f$, we match our solutions with the asymptotic solutions given below,
\begin{equation}
\label{eq:asymp}
 \lim_{r\to \text{large}} \Psil(r) = \Jl(kr)-\Nl(kr)\Kl\,.
\end{equation}
Here $\Kl$ is the reaction matrix and 
\begin{equation}\label{eq:bessel}
\begin{array}{rcl}
 [\Jl(kr)]_{ab} &=& +k_ar\,j_\ell(k_ar)\delta_{ab},\quad {\rm above\ threshold},\\
  &=& +k_ar\,\iota_\ell(k_ar)\delta_{ab},\quad {\rm below\ threshold},\\~\\
 {[\Nl(kr)]}_{ab} &=& -k_ar\,n_\ell(k_ar)\delta_{ab},\quad {\rm above\ threshold},\\
  &=& -k_ar\,\kappa_\ell(k_ar)\delta_{ab},\quad {\rm below\ threshold}.
\end{array}
\end{equation}
Here $j_\ell(x)$ and $n_\ell(x)$ denote spherical Bessel functions of first and second kinds, and $\iota_\ell(x)$ and $\kappa_\ell(x)$ are the modified spherical Bessel functions of first and second kinds, respectively. These two types of functions serve as the asymptotic forms of the wavefunction as indicated above. In the below threshold case the boundary conditions need to be changed for the excited state. In Ref.\,\cite{JOHNSON1973445, PhysRevA.32.1241}, the author has shown that in the below threshold case, only the open-open part (the part which consists of only the open channels) of the $\Kl$ matrix contributes to the final scattering matrix though one has to solve the full system of Schr\"odinger equation. In this case the asymptotic wavefunctions are either exponentially growing or decaying which may cause trouble in the numerical computation (see the second line in Eq.(\ref{eq:bessel})). It is solved by normalizing the closed channel wavefunctions and their derivatives by $\Jl$ and $\Nl$ respectively such that the new asymptotic wavefunctions become $\Jl(kr) \to 1,\ \Jl'(kr) \to \Jl'(kr)/\Jl(kr)$ and similarly for $\Nl(kr)$. 

We solve for $\Kl$ from Eq.(\ref{eq:asymp}) by taking logarithmic derivative of the equation,
\begin{equation}
\begin{array}{rcl}
 \Kl &=& [\Bl(kr_f)\Nl(kr_f)-\Nl'(kr_f)]^{-1}\\
 &&\times[\Bl(kr_f)\Jl(kr_f)-\Jl'(kr_f)]\,,
\end{array}
\end{equation}
where $\Bl(r) = \Psil'(r)[\Psil(r)]^{-1}$. Everywhere prime denotes derivative w.r.t. $r$. Once the $\Kl$ matrix is obtained, the $S$-matrix can computed through
\begin{equation}
 \Sl \equiv \mathit{1}-\Tl = (\mathit{1}+i\Kl)^{-1}(\mathit{1}-i\Kl)\,.
\end{equation}
This $\Sl$ is computed for all partial waves starting from $\ell = 0$ to $\ell_{\rm max}$. As stated in the text, the value of $\ell_{\rm max}$ depends on the initial momentum of the particles and the range of the potential. A useful lower bound on its value can be given as $\ell_{\rm max}\geq k/m_\rho$ for the case discussed in this paper. The final total cross-section is given by
\begin{equation}\label{eq:totalcs}
\begin{array}{rcl}
 [\sigma_{\rm tot}]_{ab} &=& \displaystyle\int d\Omega\dfrac{d\sigma_{ab}}{d\Omega}\\
 &=& \dfrac{1}{2k^2_b}\displaystyle\int d\Omega \left|\sum_\ell\frac{1}{2} (2\ell+1)(\tilde{\Tl})_{ab} P_\ell(\cos\theta)\right|^2\\
 &=& \dfrac{\pi}{2k^2_b}\dsum_\ell (2\ell+1)|(\tilde{\Tl})_{ab}|^2\,.
\end{array}
\end{equation}
where $(\tilde{\Tl})_{ab} = (\Tl)_{ab}+(-1)^\ell(\Tl)_{a'b}$\,. Here the prime on $a$ denotes an exchange of particles in the final 2-body DM state. Note that the last term in Eq.(\ref{eq:totalcs}) is present only when the final state particles are identical. In case of distinguishable particles, this term will be absent and so will be the extra factor of 1/2. The other two quantities of interest are the transfer and viscosity cross-sections. The definition of the transfer cross-section $\sigma_{\rm T}$ is given in Eq.(\ref{eq:transcs}). Expanding the differential cross section in the partial wave basis gives
\begin{equation}
\begin{array}{rcl}
 [\sigma_{\rm T}]_{ab} &=&\displaystyle\int d\Omega\dfrac{d\sigma_{ab}}{d\Omega}(1-\cos\theta)\\
 &=&\dfrac{\pi}{2k^2_b} \displaystyle \sum_{\ell}(\ell+1)|(\tilde{\mathcal{T}}_{\ell+1})_{ab}-(\tilde{\Tl})_{ab}|^2\,.
\end{array}
\end{equation}
Similarly the viscosity cross-section in Eq.(\ref{eq:viscositycs}) is given by
\begin{equation}
\begin{array}{rcl}
 [\sigma_{\rm V}]_{ab} &=& \displaystyle\int d\Omega\dfrac{d\sigma_{ab}}{d\Omega}\sin^2\theta\\
 &=&\dfrac{\pi}{2k^2_b}\dsum_\ell \frac{(\ell+1)(\ell+2)}{(2\ell+3)}|(\tilde{\mathcal{T}}_{\ell+2})_{ab}-(\tilde{\Tl})_{ab}|^2\,.
 \end{array}
\end{equation}

\bibliographystyle{apsrev4-1}
\bibliography{scattering.bib}
\end{document}